\documentclass[11pt,onecolumn,amssymb,nofootinbib]{revtex4}
\usepackage{amsmath, amsthm, amscd, amssymb}
\usepackage{graphicx, braket}
\usepackage{bm}
\usepackage{bbm}
\usepackage{epstopdf}

\begin{document}

\title{\bf Dynamical gravastars may evade no-go results for exotic compact objects, together with further analytical and numerical results for the dynamical gravastar model} \bigskip

\author{Stephen L. Adler}
\email{adler@ias.edu} \affiliation{Institute for Advanced Study,
Einstein Drive, Princeton, NJ 08540, USA.}

\begin{abstract}
Using graphs plotted from the Mathematica notebooks posted with our paper  ``Dynamical Gravastars'', we show that a dynamical gravastar has no hard surface, and that a second light sphere resides in the deep interior where there is maximum time dilation.  These facts may permit dynamical gravastars to evade no-go results for exotic compact objects  relating to light leakage inside the shadow, and  nonlinear instabilities arising from an interior light sphere. Testing  these surmises will require further detailed modeling calculations, beyond what we commence in this paper,  using the numerical dynamical gravastar solution.    We also discuss the effect of replacing the sigmoidal function in the gravastar calculation by a unit step function, and we analyze  why the dynamical gravastar evades the singularities predicted by the Penrose and Hawking singularity theorems, despite satisfying both the null and strong energy conditions.  We then give a simplified two-step process for tuning the initial value $\nu(0)={\rm nuinit}$ to achieve $\nu(\infty)=0$, and give exact integrals for the pressure differential equation in terms of $\nu(r)$ in the interior and exterior regions. Finally, we briefly discuss an extension of the model that includes an external shell of massive particles.

\end{abstract}

\maketitle

\section{Introduction}
The EHT observations \cite{eht} of Sg A* and M87 confirm the presence of the basic exterior spacetime geometry expected for a black hole, but leave open the question of what lies inside the light sphere.  Is it  a true mathematical black hole, or a novel  type of relativistic star  or ``exotic compact object'' (ECO),  that appears black-hole like from the outside, but has no horizon or interior singularity? The early ECO   literature has been recently  reviewed by Cardoso and Pani \cite{pani}.  In particular, they note the seminal ``gravastar'' papers of Mazur and Mattola \cite{mazur}, which are based on assuming a pressure jump in the interior equation of state, from a normal matter equation of state to the ``gravity vacuum'' equation of state proposed by Gliner \cite{gliner}, in which the pressure $p$ is minus the density $\rho$. This is the vacuum equation of state associated with a pure cosmological constant or de Sitter universe, and gives rise to the name {\bf gravastar} = {\bf gra}vity {\bf va}cuum {\bf star}. Related ideas were discussed via a condensed matter analogy in \cite{other}, \cite{other2}, \cite{khlopov1}, \cite{khlopov2}.

In a recent paper \cite{adler}  we have presented  a theory of ``dynamical gravastars'', based on following the conventional analysis \cite{zeld} of relativistic stars.    Our model differs from that of Mazur and Mottola and the subsequent paper of Visser and Wiltshire \cite{visser} in
several significant respects.  First, we perform our entire analysis from the Tolman-Oppenheimer-Volkoff (TOV) equations for relativistic stellar structure.  Second, we note that the TOV equations require that the pressure $p$ must be continuous, whereas the energy density $\rho$ can have discontinuous jumps, so we implement the Gliner equation of state by a jump to negative energy density with positive pressure.\footnote{For earlier work on gravastars with continuous pressure, but also continuous equation of state, see \cite{bened}.}  This of course violates some (but as we shall see, not all) of the classical energy conditions; however from a semiclassical quantum matter point of view, the regularized energy density is known {\it not} to obey positivity conditions \cite{Wald}, \cite{Visser1} .  Third, we avoid assuming designated radii at which transitions take place. In our model, transitions follow dynamically from the equations of motion and the assumed equations of state, hence the  terminology ``dynamical gravastar''.

The TOV equations  take the form
\vfill\eject
\begin{align}\label{newTOV}
\frac{dm(r)}{d r}=&4\pi r^2\rho(r) ~~~,\cr
\frac{d\nu(r)}{d r}=&\frac{N_\nu(r)}{1-2m(r)/r}~~~,\cr
\frac{dp(r)}{d r}=&-\frac{\rho(r)+p(r)}{2} \frac{d\nu(r)}{d r}~~~,\cr
N_\nu(r)=&(2/r^2)\big(m(r)+4\pi r^3 p(r)\big)~~~,\cr
\end{align}
where $p$ and $\rho$ are the total pressure and density, including both matter and possible cosmological constant contributions, and $\nu=\log(g_{00})$.   These equations become a closed system when  supplemented by an equation of state $\rho(p)$ giving the density in terms of the pressure. The TOV equations assign a special role to an equation of state where $p+\rho$ is zero or very small, without invoking a cosmological vacuum analogy, since $p+\rho\simeq 0$ implies a regime where the  pressure evolution with radius vanishes or is very small.  This can play a role in explaining why astrophysical black holes of very different masses have basically similar forms.

In the dynamical gravastar model, we assume an equation of state with continuous pressure $p\geq 0$,  and a jump in the density $\rho$ from an external relativistic matter state with $\rho=3p$  to an interior state with $p + \rho =\beta$, where $0<\beta<< 1 $.  We presented results in \cite{adler} and in online supplementary material in the form of Mathematica notebooks \cite{gitlab} for the cases $\beta=.1,\,.01,\,.001$, respectively labeled TOV.1, TOV.01, TOV.001.
In \cite{adler}, we included in the analysis a cosmological constant contribution, but concluded that it had a very small effect on the results of the notebooks, and so could be dropped in studies relating to  astrophysical applications.  Consequently, we omitted the cosmological constant terms in $p, \,\rho$ in an  annotated notebook which is posted online on the Wolfram Community \cite{wolf}

Our purpose here is to present additional plots obtained from the notebooks TOV.01 and TOV.001 to address objections that have been raised to interpreting the EHT observations as indicating anything other than a true black hole.  The analyses in the following sections show that these objections may be evaded by the internal structure of dynamical gravastars.  Further confirmation will require detailed simulations and computations beyond what can be inferred from the Mathematica notebooks \cite{gitlab} alone.  We give in the subsequent sections additional observations about the dynamical gravastar equations and numerical methods that we found after posting \cite{adler}, several of which will simplify and expedite further gravastar calculations.

\section{Light leakage inside the shadow}

The first objection that has been raised against an exotic compact object mimicking the galactic center black hole Sg A*, or the extragalactic hole M87,  concerns the dark space within the imaged ring, i.e., the lack of observed emission inside  the shadow.    If a postulated exotic object has a surface, then as suggested in  \cite{yuan}, reviewed in \cite{narayan},  and simulated in detail in the EHT analysis paper VI \cite{eht6},  the energy of a hot accretion inward flow striking the surface will be thermalized, giving a surface luminosity as viewed from large distances that would violate bounds set by the EHT observations.

However, this argument does not directly apply to the dynamical gravastar analyzed in \cite{adler}.  In Fig. 1 we plot the density $\rho(r)$ from the TOV.01 notebook versus radius $r$, and in Fig. 2 we give the similar plot from the TOV.001 notebook.  In both cases we see that there is no sharply defined surface at which the density jumps to its maximum value.  Instead, the density increases smoothly as $r$ decreases from the vicinity of the nominal boundary, approaching its maximum value in both cases at a radius of about 0.82 times the nominal boundary radius.  Thus, energy from an accretion flow will be dissipated over a range of radii, and the thermalization and resulting external luminosity may be significantly less than when calculated assuming a sharp surface.  Detailed simulations based on the dynamical gravastar density profile will be needed to assess whether the objections raised in the case of a sharp surface still apply.

\section{Collapse resulting from an interior second light sphere}

The second objection that has been raised against an exotic compact object mimicking  Sg A* or M87  concerns the existence of a second, interior light sphere and possible associated nonlinear instabilities.  The argument, developed in the papers \cite{cunha1}-\cite{cunha3}, \cite{ghosh}  and \cite{cardoso}, \cite{hod},  shows that on topological grounds one in general expects an even number of light spheres around a spherically symmetric exotic compact object. The outer one is the usual one, which is unstable in the sense that light near  this sphere moves away from it, either inwards or outwards.  But the general topological argument shows that there is also an inner light sphere which is stable in the sense that light moves towards it, from inside and outside.  This leads to possible nonlinear instabilities of the exotic compact object, associated with the nonlinear growth of null geodesic modes.  If the relevant time scale for instability growth is not cosmological in magnitude, this can lead to collapse or explosion of the object on observable time scales.

Writing the metric as
\begin{equation}\label{gen}
ds^2= B(r)dt^2-A(r)dr^2 - r^2(d\theta^2 + \sin^2\theta d\phi^2)~~~,
\end{equation}
the photon sphere radius is determined \cite{virb} by solution(s) of the equation
\begin{equation}\label{rph2}
\frac{d}{dr} \left(\frac{r^2}{B(r)}\right)=0~~~,
\end{equation}
which can be expanded as
\begin{equation}\label{rph3}
Q(r)\equiv 2-\frac{r}{B(r)} \frac{dB(r)}{dr}=0~~~.
\end{equation}
Writing $B(r)= e^{\nu(r)}$ as in \cite{adler}, this becomes
\begin{equation}\label{rph4}
Q(r)\equiv 2-r \frac{d\nu(r)}{dr}=0~~~.
\end{equation}
This can be rewritten by substituting  the TOV equations Eq. \eqref{newTOV}, giving after algebraic rearrangement
\begin{equation}\label{Q2}
Q(r)=3-\frac{1+8 \pi r^2 p(r)} {1- 2 m(r)/r}~~~.
\end{equation}

Since $r^2 p(r)$ and $m(r)/r$ both vanish at $r=0$ and at $r=\infty$, one has $Q(0)=Q(\infty)=2$.  This implies that $Q(r)$ must have an even number of zeros (this is the radial version of the more general topological argument of \cite{cunha1},\,\cite{cunha2}) and so in addition to the usual light sphere at $r\simeq 3M$ there must be a second light sphere.  In Fig. 3 we plot $Q(r)$ as calculated in the TOV.01 notebook, and we see that in addition to the external light sphere at $r\simeq 49.5$ there is a second zero crossing of $Q(r)$, indicating another light sphere, at the interior point $r\simeq 3.5$.  A similar plot of $Q(r)$ from the TOV.001 notebook is given in Fig. 7, in which the exterior light sphere is far off scale to the right, and the second zero crossing can be seen at $r\simeq 11.5$.

To assess the stability of the interior light sphere, we follow the analysis of \cite{cardoso}, which shows that stability (instability) corresponds to a negative (positive) value of the second derivative of the potential $V^{\prime \prime}$, which \big(omitting  a positive factor $L^2/\big(A(r) r^4\big) $, with $L$ the angular momentum\big)  is given by
\begin{equation}\label{vsec}
V^{\prime \prime}(r)=2-r^2 B^{\prime \prime}(r)/B(r)~~~.
\end{equation}
This can be rewritten in terms of quantities appearing in the TOV equations as
\begin{align}\label{vsec1}
V^{\prime \prime}(r)=& 2-r^2 (\nu^{\prime \prime}(r)+\nu^\prime(r)^2)~~~,\cr
\nu^{\prime}(r)=&\frac{N_\nu(r)}{(1-2m(r)/r)}~~~,\cr
dN_\nu(r)/dr=&-(4/r^3)\big(m(r)+4 \pi r^3 p(r)\big) +(2/r^2)\big(dm(r)/dr+ 12 \pi r^2 p(r)+4\pi r^3 dp(r)/dr \big)~~~,\cr
\nu^{\prime \prime}(r)=&\frac{dN_\nu(r)/dr}{1-2m(r)/r}+\frac{2N_\nu(r)\big(r^{-1}dm(r)/dr-m(r)/r^2\big)}{(1-2m(r)/r)^2}~~~.\cr
\end{align}
In Fig. 4 we plot $V^{\prime \prime}(r)$ for the TOV.01 notebook, showing that it is positive at at the outer light sphere radius of $r\simeq 49.5$, indicating instability,  and may be negative at the inner light sphere radius of $r\simeq 3.5$.  In Fig. 5 we repeat this plot with a much finer vertical scale, showing that $V^{\prime \prime}(r)$ is negative, indicating stability, at the inner light sphere radius of $r\simeq 3.5$.  In Fig. 8, we plot $V^{\prime \prime}(r)$ for the TOV.001 notebook, showing that it is negative, again indicating stability, at the inner light sphere radius $r\simeq 11.5$.

Stability of the inner light sphere raises the possibility of a pileup of null geodesics at that radius, leading to a possible dynamical instability that, on a sufficiently long time scale, could blow up or collapse an exotic compact object \cite{keir}--\cite{benomio}, \cite{cunha3}.  Simulations for two models of bosonic compact objects in \cite{cunha3} suggests that for these models this instability occurs on physically accessible time scales, ruling out these compact objects as candidates for black hole mimickers. However, for dynamical gravastars the situation can differ, and this objection may be evaded. In Fig. 6 we plot $\nu(r)= \log B(r) = \log g_{00}(r)$ for the TOV.01 notebook, which shows that the inner light sphere radius corresponds to an exponentially small $\nu(r)$, and therefore an exponentially large time dilation.  In Fig. 9 we give a similar plot for the TOV.001 notebook, showing that the smallness of $\nu(r)$ at the inner light sphere radius is even more extreme, and the trend shows that as $\beta$ approaches zero, the trend of $\nu(r)$ at the inner light sphere radius is to even smaller values than shown in Figs. 6 and 9.  This means that for parameters giving physically realistic gravastars, as discussed further in Sec. 8, the time scale for instability development at the inner light sphere can be very large on a cosmological time scale.  Detailed dynamical gravastar simulations will be needed to assess whether the objections raised in \cite{cunha3} are relevant.

\section{Results when a sigmoidal jump is replaced by a unit step jump}

The calculations of the previous sections based on the Mathematica notebooks associated with \cite{adler} were all done with a density jump smoothed by a sigmoidal function
\begin{equation}\label{smooththeta}
\theta_\epsilon(x)=\frac{1}{1+e^{-x/\epsilon}}~~~,\
\end{equation}
with $\epsilon=.001$. By changing one line of Mathematica code, $\theta_\epsilon(x)$ can be replaced by a Heaviside step function $\theta(x)$, represented  by the Mathematica function UnitStep[x].  When this is done in the TOV.01 notebook, the initial value of the metric exponent $\nu(x)$, denoted by nuinit in the program, has to be retuned from nuinit = -21.255 to
nuinit = -23.628 in order to achieve the boundary condition $\nu(\infty)=0$.   This results in a substantial change in the nominal boundary from  $2M\simeq 33$   to  $2M\simeq 60$, but the qualitative features of the solution are not altered.  In
Figures 10, 11, and 12 we give the results obtained in the unit step jump case that are analogous to those shown in Figures 1, 3, and 5  obtained respectively from the sigmoidal function case.  The conclusions reached above are unchanged.  Similar results to those for the unit step are obtained from a sigmoidal density jump when one  takes $\epsilon=.00001$ in place of $\epsilon=.001$.

\section{Energy conditions and black hole singularity theorems.  }

The gravastar solutions developed in \cite{adler} do not have central singularities.  So why do the classic singularity theorems of general relativity fail to apply?  One  might think that this is because of the negative interior energy density $\rho <0$ which violates both the ``weak energy condition'' $\rho \geq 0,\, \rho+p \geq 0$ and the ``dominant energy condition'' $\rho \geq |p|$.  However, the seminal Penrose singularity theorem \cite{penrose} uses the ``null energy condition''
\begin{equation}\label{nullen}
\rho+p \geq 0~~~,
\end{equation}
which is obeyed by the assumption $\rho+p=\beta>0$, and the subsequent Hawking theorem \cite{hawking} uses the ``strong energy condition''
\begin{equation}\label{stringen}
\rho+p\geq 0,\, \rho+3p \geq 0~~~,
\end{equation}
which is also obeyed by $\rho+p=\beta >0,\,p>0$.  So the failure of these theorems to apply is not a result of the breakdown of an energy condition assumption.

However, as emphasized in the retrospective \cite{senovilla} on the Penrose theorem and its successors, an ``initial/boundary condition is absolutely essential in the theorems''.  For example, the Penrose theorem assumes the presence of a closed, future-trapped surface.  But because the dynamical gravastar solution has $B(r) > 0$ and $A(r)^{-1}=1-2m(r)/r > 0$, as shown in the graphs of \cite{adler}, there is no trapped surface, corresponding to the fact that there is neither an event horizon nor an apparent horizon \cite{adler2}.  Hence despite obeying the null energy condition, the dynamical gravastar does not satisfy the conditions needed for the Penrose singularity theorem. Similarly, the Hawking theorem requires a negative extrinsic curvature relative to the positive time axis over some spacelike surface, and this is not present in the gravastar because there are no horizons.   Thus despite obeying the strong energy condition, the dynamical gravastar does not satisfy the conditions needed for the Hawking singularity theorem.  We believe these results are significant, because the strong and null energy conditions are the natural relativistic generalization of the intuitive nonrelativistic notion that matter should have a nonnegative energy density.

The null and strong energy conditions have a direct consequence for the TOV equations in the central neighborhood  $r \simeq 0$ when the density  $\rho(r)$ is approximately constant near the origin, as is the case for our gravastar models.  When $\rho(r) \simeq \rho(0)$ for small $r$, integrating the differential equation for $m(r)$ gives
\begin{equation}\label{minteg}
m(r) \simeq \frac{4\pi r^2}{3}\rho(0) \simeq  \frac{4\pi r^2}{3}\rho(r) ~~~.
\end{equation}
Equation \eqref{newTOV} then gives for $N_\nu(r)$
\begin{equation}\label{Nneweq}
N_\nu(r)\simeq  \frac{4\pi r^2}{3} \big(\rho(r) +3 p(r) \big)~~~,
\end{equation}
which is nonnegative by virtue of the strong energy condition.  This implies that near $r=0$, the derivative $d\nu(r)/d r$ is nonnegative, since the denominator $1-2m(r)/r$ is very close to unity.  Hence $\nu(r)$ starts from the origin as a nondecreasing function of radius $r$.  The TOV equation for $dp(r)/d r$ then implies, by virtue of both the null and strong energy conditions, that $p(r)$ starts from the origin as a nonincreasing function of radius $r$.  This behavior is seen in our numerical solutions, where $\nu(r)$ is a monotonic increasing and $p(r)$ is a monotonic decreasing function of $r$.

\section{A simple two-step method for tuning $\nu(0)={\rm nuinit}$}

As noted in \cite{adler} and in Sec. 1, the effect of including a cosmological constant in the calculation is extremely small, so for astrophysical applications of dynamical gravastars it can be dropped.  The TOV equations then take the form given in Eq. \eqref{newTOV},
where $p$ and $\rho$ are now the matter pressure and density.

We see that in Eqs. \eqref{newTOV}, $\nu(r)$ enters only through its derivative $d\nu(r)/dr$.   Hence a constant shift in $\nu(r)$, such as changing the initial value $\nu(0)$,  only affects $\nu(r)$ itself; the equations and numerical results for $m(r)$ and $p(r)$ are unaffected.  Thus permits a simple method for tuning the initial value $\nu(0)$ to achieve the large $r$ boundary condition $\nu({\rm rmax})=0$, where we have taken rmax as a proxy for $r=\infty$.

As noted in Fig. 9 of \cite{adler}, when nuinit is correctly tuned, exterior to the nominal horizon at $2M$ the metric coefficient $g_{00}=e^{\nu(r)}$ almost exactly coincides with the Schwarzschild metric value $g_{00}=1-2M/r$, with $M$ the gravastar mass.  This is a consequence of the fact that the density $\rho(r)$ becomes very small beyond $2M$, so by the Birkhoff uniqueness theorem for matter-free spherically symmetric solutions of the Einstein equations, the metric must take the Schwarzschild form.  A consequence of this, and of the TOV equation properties under shifts in nuinit, is that with an arbitrary initial guess for nuinit, the large $r$ value of $e^{\nu(r)}$ must have the form
\begin{equation}\label{largeR}
e^{\nu(r)}=e^K (1-2M/r)~~~,
\end{equation}
where $K$ is a constant and the gravastar mass $M$ can be read off from the exterior value of $m(r)$, since $m(r)$ is not affected by shifts in nuinit.  Thus, reducing the initial guess for nuinit by $K$, with
\begin{equation}\label{Kvalue}
K=\nu(r)-\log(1-2M/r) \simeq \nu({\rm rmax})-\log\big(1-2 m({\rm rmax})/{\rm rmax}\big)
\end{equation}
gives the correct tuning of nuinit.  So the two step procedure to tune nuinit is: (i)  first run the program with an initial guess for nuinit, and calculate $K$ from Eq. \eqref{Kvalue}, (ii)  then  subtract this from the initial guess for nuinit and rerun the program.  We found that this worked well in practice, with residual values of $K$ after the second step of order $10^{-9}$.

\section{Integration of the equation for $p(r)$ in terms of $\nu(r)$}

Referring again to the  TOV equations given in Eq. \eqref{newTOV}, we find that for the matter equations of state assumed in \cite{adler}, and zero cosmological constant,  the differential equation for $p(r)$ can be integrated in closed form in terms of $\nu(r)$.  For pressures $p>{\rm pjump}$, where the equation of state is assumed to take the form
$p(r)+\rho(r)=\beta$, the differential equation for $p(r)$ takes the simple form
\begin{equation}\label{pabove}
\frac{dp(r)}{d r}=-\frac{\beta}{2} \frac{d\nu(r)}{d r}~~~,
\end{equation}
which can be integrated, using $p(0)=1$ and $\nu(0)={\rm nuinit}$, to give
\begin{equation}\label{pabove1}
p(r)=1 + \frac{\beta}{2} \big( {\rm nuinit}-\nu(r)\big)~~~.
\end{equation}
For pressures $p<{\rm pjump}$, where the equation of state is assumed to take the form
$\rho(r)=3 p(r)$, the differential equation for $p(r)$ also takes a simple form
\begin{equation}\label{pbelow}
\frac{dp(r)}{d r}=-2 p(r) \frac{d\nu(r)}{d r}~~~,
\end{equation}
which can be integrated to give
\begin{equation}\label{pbelow1}
p(r)={\rm pjump}\, e^{2 \big(\nu( {\rm rjump})-\nu(r)\big)}~~~,
\end{equation}
where rjump is the radius value at which the density jump is  dynamically determined to occur by the TOV equations.  For example, for the plots shown in Figs. 10  -- 12, the density jump at pjump=.95 is found to occur at rjump=48.895.
We have verified that plots of $p(r)$ using these exact solutions  agree with plots of $p(r)$  obtained from numerical integration of the TOV equations.

\section{Parameter Values and Quantitative Results for Time Delays}

In this section we present further computational results obtained from the notebooks TOV.01 and TOV.001, with the cosmological constant set equal to zero, permitting the simple two-step procedure for tuning the initial value $\nu(0)$ discussed in Sec. 6.  We retain the sigmoidal smoothed step function used in these notebooks, since this has no bearing on the initialization procedure. We also continue to assume that the pressure at the origin of coordinates is initialized to $p(0)=1$, which is  always possible by rescaling the radial coordinate and is discussed in detail in \cite{adler}.
This leaves two essential physical parameters of the dynamical gravastar model, the numerically small constant $\beta$ in the interior equation of state $p+\rho=\beta$, and the pressure value pjump at which the equation of state switches to the exterior relativistic matter equation of state $\rho=3p$.

What are reasonable physical ranges for these two parameters?  Presumably, the central pressure $p(0)=1$ corresponds to a Planck energy-scale pressure in our geometrized units, and then the value of pjump is the fractional value of this pressure at which the equation of state jumps.  Since all the values of pjump for which we were able to obtain numerical results were  above $.85$ for the TOV.01 notebook, and above $.9773$ for the TOV.001 notebook, these also correspond to pressures within a decade  of a Planck energy-scale pressure.  The possible range for the parameter $\beta$ is potentially much larger, and less clear.    Since pressure p and energy density $\rho$ scale with length as the inverse of length squared, corrections to the  equation of state of hypothesized quantum matter in the Planckian regime could come from anywhere between just below the Planck scale, corresponding to $\beta \sim .01$, to the electroweak symmetry breaking scale where particles get their masses, corresponding to $\beta \sim (10 ^{-17})^2=10^{-34}$.  In practice, we could not get the Mathematica notebooks to perform well at $\beta$ values much lower than  the value $\beta=.001$ computed in the TOV.001 notebook.

Because of the radial rescaling covariance of the model, only ratios of quantities with the same dimensionality have a physical significance.  Since in geometrized units the velocity of light is unity, the coordinate time delay $\Delta T$ for a particle to traverse from the exterior to the center of a  gravastar, and the corresponding coordinate radius of the effective ``horizon'' $R = 2 M$, with $M$ the effective gravastar mass, both have dimensions of length.  So the ratio (the normalized time delay)
\begin{equation}\label{ratioone}
D \equiv \frac{\Delta T}{2 M}
\end{equation}
has a physical significance, and we proceed to compute tables of $D$ versus pjump for  $\beta=.01$ and $\beta=.001$, using the notebooks TOV.01 and TOV.001 respectively.  As in Sec. 6, the gravastar mass $M$ is read off from the exterior value of $m(r)$, clearly visible in the plots as the value where the $m(r)$ curve levels off.

To calculate the coordinate  time delay $\Delta T$, we use the text of  Weinberg \cite{weinberg}, which conveniently computes test particle motion in the general spherically metric of Eq. \eqref{gen}.  The radial part of the equation of motion is given by
\begin{equation}\label{radial}
\frac{A(r)}{B^2(r)} \left(\frac{dr}{dt}\right)^2 +\frac{J^2}{r^2}-\frac{1}{B(r)}=-E~~~,
\end{equation}
with the constants of motion energy per unit mass $E$ and angular momentum per unit mass $J$ given in terms of the test particle velocity $V$ at infinity and impact parameter $b$ by
\begin{equation}\label{constants}
J=bV~~~,\, E=1-V^2~~~.
\end{equation}
From these we get
\begin{align}\label{DeltaT}
\Delta T=&\int_{r^*}^R dr \frac{dt}{dr}~~~\cr
=&\int_{r^*}^R  dr \left( \frac{A(r)}{B(r)} \right)^{1/2} /[1-B(r)(1-V^2+b^2V^2/r^2)]^{1/2}~~~,\cr
\end{align}
with $r^*$ the inner radius at which the denominator  in Eq. \eqref{DeltaT} vanishes, and with $B(r)$ and $A(r)$ given in terms of quantities appearing in the TOV equations of Eq. \eqref{newTOV} by
\begin{align}\label{AandB}
B(r)=& e^{\nu(r)}~~~,\cr
A(r)=&1/[1-2m(r)/r]~~~.\cr
\end{align}

In using Eq. \eqref{DeltaT} in computations with the TOV.01 and TOV.001 notebooks, it turns out that $B(r)$ is so small in the gravastar interior that the denominator $[1-B(r)(1-V^2+b^2V^2/r^2)]^{1/2}$ is approximately equal to unity for relevant values of $V \leq 1$ and $0 \leq b/(2M)  \leq 1$.  Hence to one tenth of one percent accuracy or better for all entries in the Tables below, the formula for $\Delta T$ can be simplified to\footnote{Similar reasoning can be used to strongly bound the change in azimuthal angle $\Delta \phi$ of an infalling object inside the gravastar.  The differential equation obeyed by $\phi$ is \cite{weinberg} $d\phi/dt = J B(r)/r^2$, and so the change with radius is  $\Delta \phi=\int dr (dt/dr) d\phi/dt$.   This leads to the bound $|\Delta \phi| \leq \int_{r_{\rm min}}^R [A(r)B(r)]^{1/2} (2M/r^2)$, which evaluated for the final line of Table II gives $|\Delta \phi| \leq 2 \times 10^{-11}$. This shows that  test bodies move inside the gravastar on nearly radial trajectories.}
\begin{equation}\label{DeltaT1}
\Delta T
=\int_{r_{\rm min}}^R  dr \left( \frac{A(r)}{B(r)} \right)^{1/2} ~~~,
\end{equation}
with $r_{\rm min}$ the inner radius cutoff of $10^{-7}$ used in the notebooks.  This simplification has the important consequence that the time delays for an infalling object are essentially independent of its impact parameter, and are the same as would be computed for the case of an incoming photon on a radial trajectory with $V=1,\,b=0$.  In Tables I and II we tabulate values of $D$, $M$, and the rmax  used in each calculation, for a range of values of pjump.  We see that as pjump decreases to the smallest value for which the notebooks give stable results, the magnitude of $D$ increases up to a maximum of $1.43  \times 10^7$ for $\beta = .01$, and $1.54 \times 10^7$ for $\beta=.001$. The fact that these limits are so similar suggests that they relate to the computational accuracy of the Mathematica utility, and do not reflect a systematic trend versus $\beta$  from which one could attempt to extrapolate to smaller  $\beta$ values.

To scale up these numbers to astrophysical black holes, we start from the light transit time across the Schwarzschild radius $2M_\odot$ of a solar mass black hole, which is $3 {\rm km}/(3 \times 10^5 {\rm km/s}) \simeq 10^{-5} {\rm s} = 0.3 \times 10^{-12} {\rm yr}$. For this size gravastar, the maximum time delay $D$ calculated in Tables I and II gives a transit time from the nominal horizon to the center of
$1.5 \times 10^7 \times  0.3 \times 10^{-12} {\rm yr} \simeq 0.5 \times 10^{-5} {\rm yr}$.  The corresponding  transit times for a $4 \times 10^6 M_\odot$ gravastar and a $10^8  M_\odot$ mass gravastar  are 20 years and 500 years respectively.  Thus the transit time delays for supermassive black holes in the gravastar model can accommodate the time delays reported by Cendes et al. \cite{cendes} for delayed radio emission following optical observation of tidal disruption events in a sampling of supermassive black holes.

From the above calculations, we get some general observations about time delays arising from the very small value of $B(r)$ in gravastars: (i) Assuming astrophysical black holes of different masses have homologous structures, differing only by a rescaling of the central pressure, then time delays scale proportionally to the mass of the black hole.  The proportionality constant is a function of the parameters $\beta$ and pjump of the gravastar model, as is clear from comparing the different lines in the tables, and so cannot be predicted a priori.  (ii) Time delays can easily range from minutes to centuries depending on the mass of the hole.  (iii) For impact parameters less than or equal to the nominal Schwarzschild radius, the time delay is independent of both impact parameter $b$ and the velocity at infinity $V$ of an infalling object.

Finally, we return to the discussion of Sec. 3 and calculate the normalized time delay over the inner light sphere, which for all pjump values in Table II has a radius of 11.5.   Defining $D_{\rm inner}$ by
\begin{equation}\label{inner}
D_{\rm inner}
=\int_{r_{\rm min}}^{11.5} dr \left( \frac{A(r)}{B(r)} \right)^{1/2}/11.5 ~~~,
\end{equation}
we find that for pjump=.9773, the lowest line in Table II, $D_{\rm inner} \simeq 10^{12}$, five orders of magnitude larger than the normalized delay from the gravastar surface to center. This qives quantitative support to the discussion given in Sec. 3 of possible evasion in the gravastar model of conjectured instabilities associated with the inner light sphere.

\begin{table} [ht]
\caption{Normalized time delay $D$ defined in Eq.\eqref{ratioone} and hole mass $M$, versus   pjump and maximum radius rmax used in the calculation, all for $\beta=.01$.}
\centering
\begin{tabular}{c c c  c}
\hline\hline
pjump&$D$&$M$ &rmax\\
\hline
.95  &  680   & 16.5    & 60    \\
.94  & 1840    &  44.4   & 200    \\
.93  & 4980    & 120    & 600    \\
.92  & 13,500    & 322    & 1,000    \\
.90  & 98,800   &  2,330   & 8,000    \\
.88  &  723,000   &16,900     & 80,000    \\
.86  & 5.29 $\times 10^6$   & 122,000    & 400,000    \\
.85  & 1.43 $\times 10^7$    & 329,000    &800,000     \\
\hline\hline
\end{tabular}
\label{tab1}
\end{table}

\begin{table} [ht]
\caption{Normalized time delay $D$ defined in Eq.\eqref{ratioone} and hole mass $M$, versus  pjump and maximum radius rmax used in the calculation, all for $\beta=.001$. }
\centering
\begin{tabular}{c c c  c}
\hline\hline
pjump&$D$&$M$ &rmax\\
\hline
 .98 &  1.04$ \times 10^6$   & 25,500    & 90,000    \\
.9795  & 1.71$\times 10^6$    & 42,000    & 200,000    \\
.9793  & 2.09$\times 10^6$    &51,300     &200,000     \\
 .9791 &2.55$\times 10^6$     &  62,600   & 200,000    \\
.9790  &2.82 $\times 10^6$    & 69,200    & 200,000   \\
.9785 &4.64$\times 10^6$       &114,000    &250,000   \\
 .9780 &7.66$\times 10^6$     &188,000     &500,000     \\
.9775  &  1.26$\times 10^7$   & 310,000    & 900,000    \\
.9773  & 1.54$\times 10^7$    & 379,000    & 900,000    \\
\hline\hline
\end{tabular}
\label{tab2}
\end{table}

\section{Extension of the model to include an external layer with nonzero particle masses}

The model of \cite{adler} assumes an exactly massless equation of state $\rho(p)=3p$ in the region external to the density jump at pressure pjump.  A more realistic model would be to included an outer shell in which particle masses are taken into account, most simply through a ``dust'' equation of state  $\rho(p)=\rho_0={\rm constant}$ for pressure $p$ below a chosen value pext.   We find that for $\rho_0={\rm pext}=10^{-8}$ or smaller, the results are very similar to the original model. Another method to approximate the effects of a matter layer is to replace the pressure $p$ in the exterior region equation of state by $(p^2+\rho_0^2)^{1/2}$, with $\rho_0=10^{-8}$ or smaller, again giving results very similar to the original model.    To get an idea of a realistic physical range for the two new  parameters, we note that the dimensions of energy density and pressure are both ${\rm length}^{-2}$ in the geometrized units we are using.  Let us assume that  the density jump at pressure pjump occurs in the regime between the grand unification scale  and the Planck scale, that is in the range $10^{16}$ to $10^{19}$ GeV, while the jump at pext $\simeq \rho_0$ occurs at the electroweak symmetry breaking scale of order 160 GeV. Then the corresponding ratio ${\rm pext}/{\rm pjump}$ will be of order $(160/10^{16})^2 \sim 10^{-28}$ or smaller, far below the level of $\sim 10^{-8}$ at which the model already shows no significant change from including a massive external layer or an equivalent modification to the way pressure enters the equation of state.

\section{Discussion}

The preceding sections give a number of results that substantially extend our initial paper \cite{adler}.  They group into three principal categories:

  The first category is an examination, in the context of dynamical gravastars, of objections that have been raised against the possibility that an exotic compact object (ECO) is what is  being seen in the EHT observations, rather than a mathematical black hole with a horizon.  Based on our gravastar Mathematica notebooks, we give graphs showing
 that a dynamical gravastar has no hard surface, and  that a second light sphere resides in the deep interior where there can be exponentially large time dilation, for which we give quantitative estimates.   These computed results are at variance with the assumptions made in the papers that claim to falsify an ECO interpetation of observed astrophysical black holes.  A definitive  statement will require a much extended investigation and computations,  but stating the motivation for such an investigation is in important initial step.

 The second category are some interesting general theoretical features of the gravastar model.  We note that the Penrose and Hawking singularity theorems are evaded in this model, even though it  accurately mimics the external geometry of a black hole, and despite the fact that the null energy condition and the strong energy condition are obeyed, because a trapped surface never forms.   We also show, through computation,  that a physically realistic external massive shell can be harmlessly added to the model, which is a significant extension of the results in our posted Mathematica notebooks.

 The third category are several simplifications of the TOV equations of the model that will streamline  further exploration.  First, we note that the model is qualitatively unchanged if the cosmological constant is set to zero, and if the smoothed sigmoidal jump is replaced by a unit step jump.  This eliminates two extraneous parameters of the original model, and thus significantly reduces the dimensionality of the model parameter space.  Additionally, when the cosmological constant is set to zero the TOV equations, as supplemented with with the matter equations of state assumed in \cite{adler}, can be partially integrated to give closed form expressions for the pressure $p(r)$ in terms of the metric exponent $\nu(r)$ (and by inversion, vice versa). And again when there is no cosmological constant, the procedure for ``tuning'' the initial value nuinit=$\nu(0)$ is greatly simplified.  We have substantiated the value of these simplifications in new work \cite{adler3} that explores the residual parameter space of the dynamical gravastar model, and also use them in the quantitative estimates of time delays given in Sec. 8 of this paper.

 \section{Acknowledgement}

 I wish to thank Brent Doherty for stimulating conversations relating to the material in Secs. 6 and 7. I also         thank the preliminary referees for useful comments, and both the final referee and George N. Wong for asking for a quantitative calculation of time delays.

\vfill\eject
\begin{figure}[t]
\begin{centering}
\includegraphics[natwidth=\textwidth,natheight=300,scale=0.8]{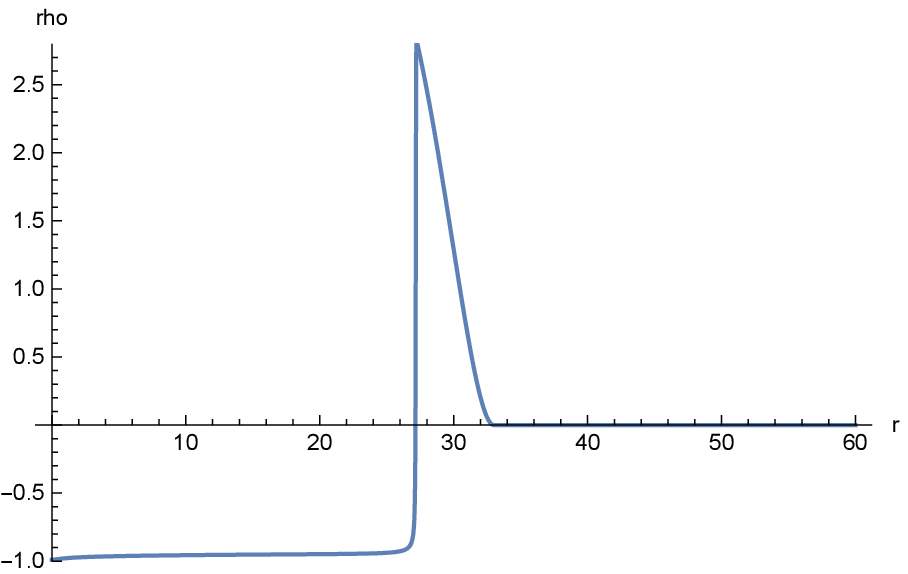}
\caption{Plot of the density $\rho(r)$ for the TOV.01 notebook. The nominal boundary is $2M\simeq 33$. }
\end{centering}
\end{figure}

\begin{figure}[t]
\begin{centering}
\includegraphics[natwidth=\textwidth,natheight=300,scale=0.8]{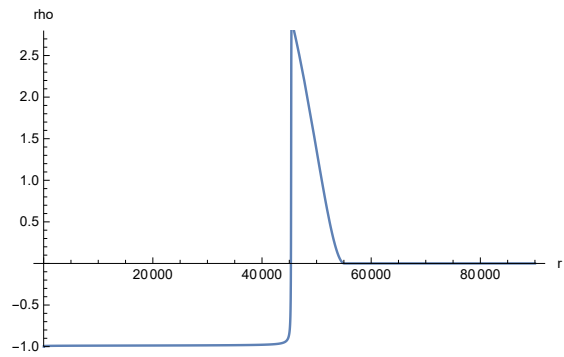}
\caption{ Plot of the density $\rho(r)$ for the TOV.001 notebook. The nominal boundary is $2M\simeq 55,200$. }
\end{centering}
\end{figure}

\begin{figure}[t]
\begin{centering}
\includegraphics[natwidth=\textwidth,natheight=300,scale=0.8]{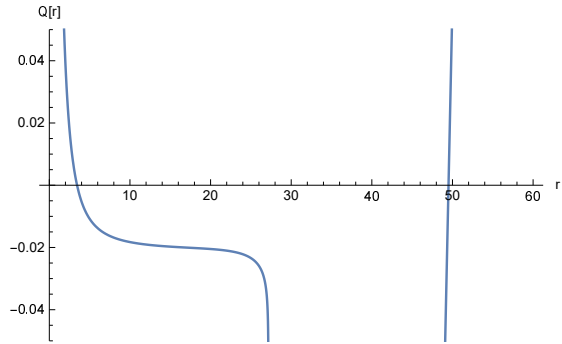}
\caption{Plot of $Q(r)$ for the TOV.01 notebook.  The exterior light sphere is the zero at $r=3M\simeq 49.5$; there is an interior second light sphere at $r\simeq 3.5$.   }
\end{centering}
\end{figure}

\begin{figure}[t]
\begin{centering}
\includegraphics[natwidth=\textwidth,natheight=300,scale=0.8]{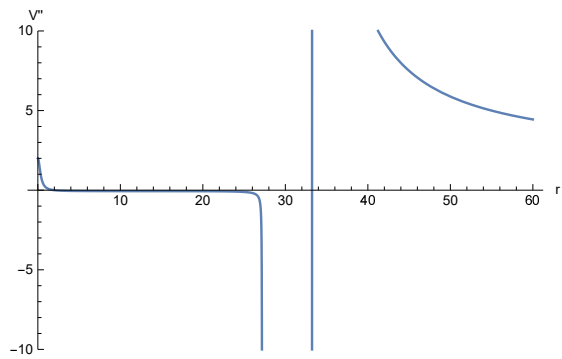}
\caption{The quantity $V^{\prime\prime}$ of Eq. (36) of Cardoso et. al \cite{cardoso2} for the TOV.01 notebook, with the positive factors $L^2/r^4$ scaled out.  One sees that  $V^{\prime\prime}$  is positive at $r\simeq 49.5$, and may be negative at $r\simeq 3.5$.  }
\end{centering}
\end{figure}

\begin{figure}[t]
\begin{centering}
\includegraphics[natwidth=\textwidth,natheight=300,scale=0.8]{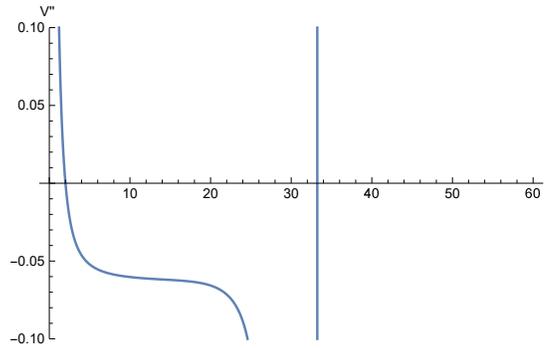}
\caption{The quantity $V^{\prime\prime}$ of Eq. (36) of Cardoso et. al \cite{cardoso2} for the TOV.01 notebook, with the positive factors $L^2/r^4$ factored out, plotted with a much finer vertical scale than used in Fig. 4.   One sees that  $V^{\prime\prime}$  is negative at $r\simeq 3.5$.    }
\end{centering}
\end{figure}

\begin{figure}[t]
\begin{centering}
\includegraphics[natwidth=\textwidth,natheight=300,scale=0.8]{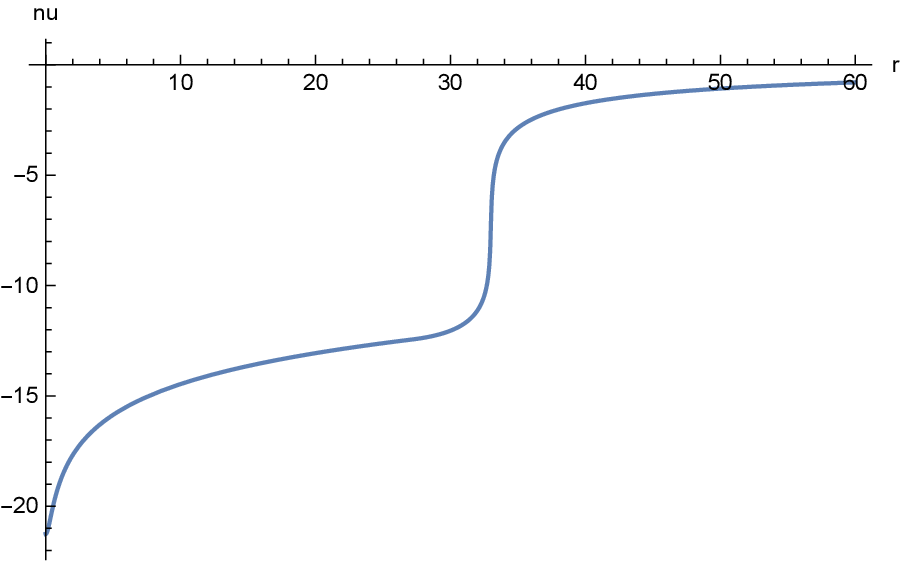}
\caption{Plot of $\nu(r)$ for the TOV.01 notebook.  The large negative value at $r\simeq 3.5$ corresponds to a large time dilation.  }
\end{centering}
\end{figure}

\begin{figure}[t]
\begin{centering}
\includegraphics[natwidth=\textwidth,natheight=300,scale=0.8]{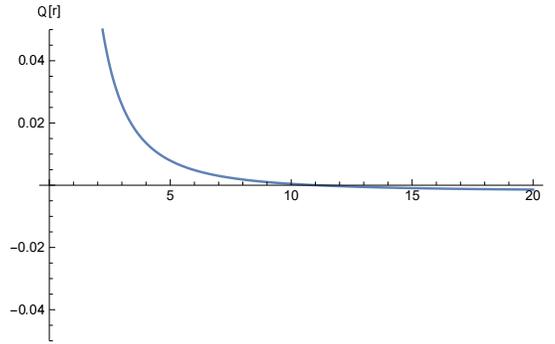}
\caption{Plot of $Q(r)$ for the TOV.001 notebook.  The exterior light sphere is a zero at $r=3M\simeq 82,800$, far off scale to the right; there is an interior second light sphere at $r\simeq 11.5$.     }
\end{centering}
\end{figure}

\begin{figure}[t]
\begin{centering}
\includegraphics[natwidth=\textwidth,natheight=300,scale=0.8]{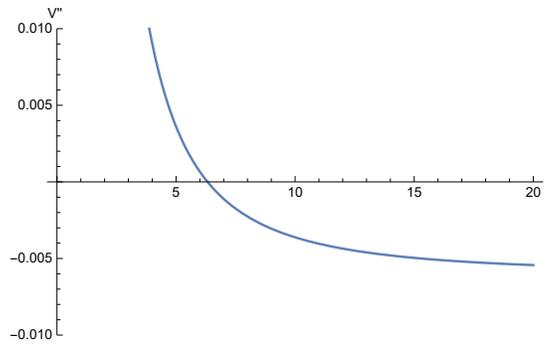}
\caption{The quantity $V^{\prime\prime}$ of Eq. (36) of Cardoso et. al \cite{cardoso2} for the TOV.001 notebook, with the positive factors $L^2/r^4$ factored out.  One sees that  $V^{\prime\prime}$  is negative at $r\simeq 11.5$.    }
\end{centering}
\end{figure}

\begin{figure}[t]
\begin{centering}
\includegraphics[natwidth=\textwidth,natheight=300,scale=0.8]{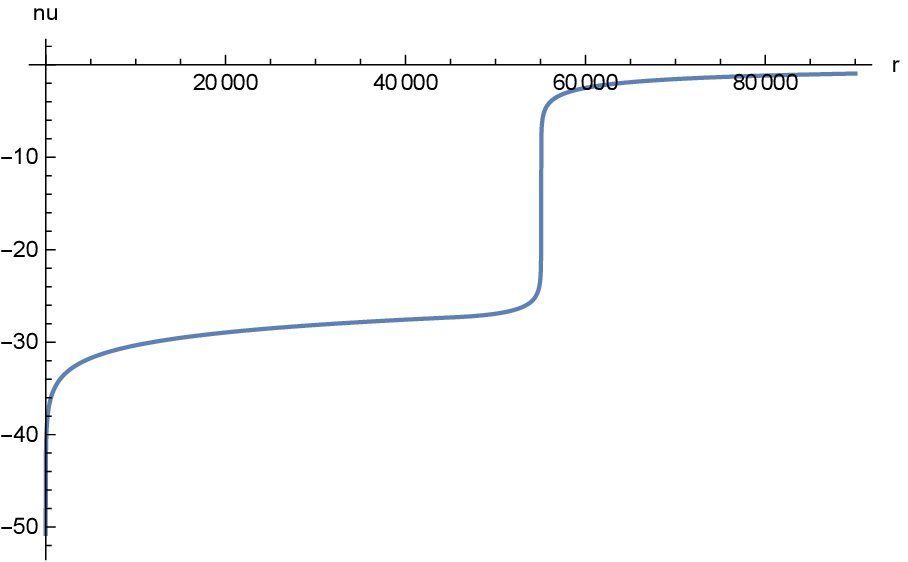}
\caption{Plot of $\nu(r)$ for the TOV.001 notebook.  The large negative value at $r\simeq 11.5$ corresponds to a large time dilation.   }
\end{centering}
\end{figure}

\vfill\eject
\begin{figure}[t]
\begin{centering}
\includegraphics[natwidth=\textwidth,natheight=300,scale=0.8]{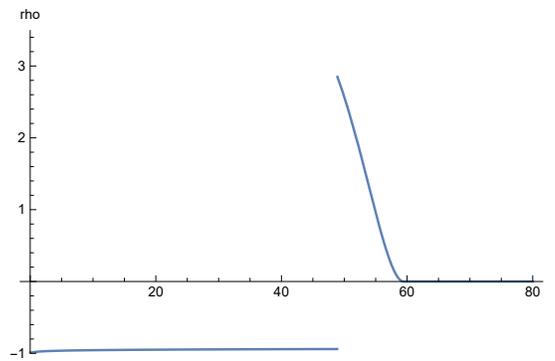}
\caption{Plot of the density $\rho(r)$ for the TOV.01 notebook when the sigmoidal jump used in Fig. 1 is  replaced by a unit step jump.  The nominal boundary is now $2M\simeq 60$. }
\end{centering}
\end{figure}

\begin{figure}[t]
\begin{centering}
\includegraphics[natwidth=\textwidth,natheight=300,scale=0.8]{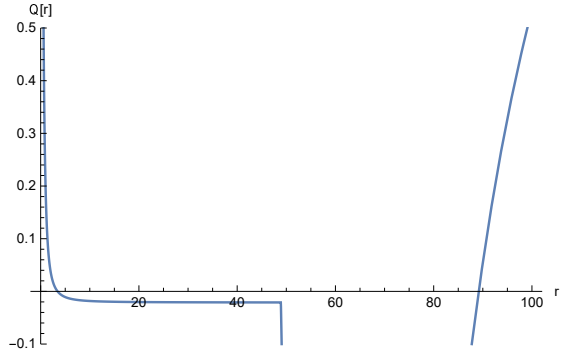}
\caption{Plot of $Q(r)$ for the TOV.01 notebook when the sigmoidal jump used in Fig. 3 is  replaced by a unit step jump.  The exterior light sphere is the zero at $r=3M\simeq 90$; there is an interior second light sphere at $r\simeq 3.7$.   }
\end{centering}
\end{figure}

\begin{figure}[t]
\begin{centering}
\includegraphics[natwidth=\textwidth,natheight=300,scale=0.8]{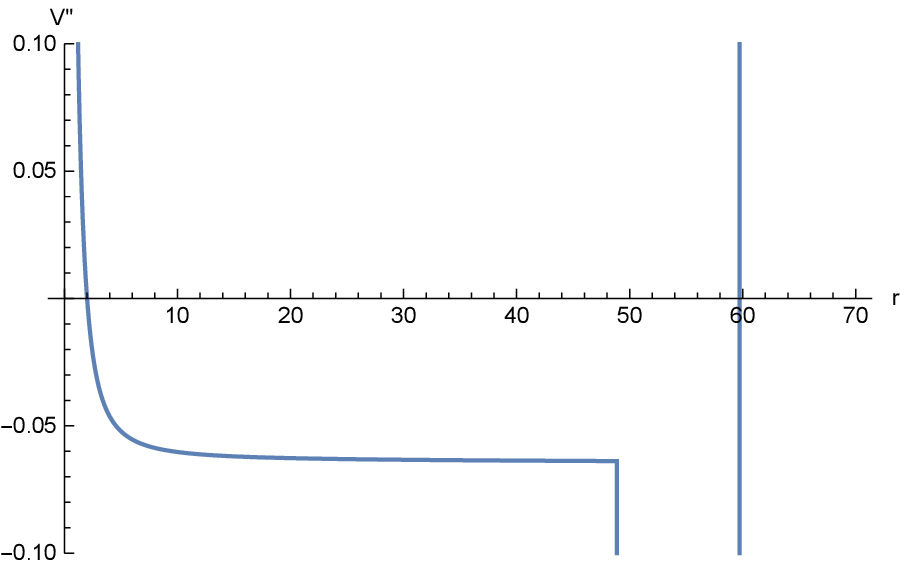}
\caption{The quantity $V^{\prime\prime}$ of Eq. (36) of Cardoso et. al \cite{cardoso2} for the TOV.01 notebook when the sigmoidal jump used in Fig. 5 is  replaced by a unit step jump, with the positive factors $L^2/r^4$ factored out.    One sees that  $V^{\prime\prime}$  is negative at $r\simeq 3.7$.    }
\end{centering}
\end{figure}


\begin{thebibliography}{99}
\bibitem{gitlab}  For active links directly accessing the Mathematica notebooks used in \cite{adler}, and also accessing a simplified and annotated notebook prepared for the Mathematica Community, see the abstract of \cite{adler}.   These notebooks were written in Mathematica version 12.2. They use the functions Exp, Log, Print, Plot, and Show, which were introduced in version 1.0; NDSolve and Evaluate, which were introduced in version 2.0; and LogPlot, which was introduced in version 6.0.
\end{thebibliography}
\end{document}